\documentclass[11pt,a4paper,oneside]{article}
\usepackage[top=1in,bottom=1in,left=.5in,right=.5in]{geometry}
\usepackage[margin=.5in,small]{caption}

\usepackage{amssymb,amsmath}
\usepackage{authblk}
\usepackage{cite}
\usepackage{amsmath,gensymb}
\usepackage{epsfig}
\usepackage{graphics}
\usepackage{amsfonts, amssymb, upgreek}
\usepackage{textcomp}
\usepackage[english]{babel}
\usepackage{blindtext}
\usepackage{color}

\providecommand{\abs}[1]{\vert #1 \vert}

\newcommand{\xiI}{\xi_\mathrm{_I}}
\newcommand{\xiL}{\xi_\mathrm{x}}
\newcommand{\xiV}{\xi_\mathrm{h}}
\newcommand{\xiT}{\xi_\mathrm{t}}
\newcommand{\nuI}{\nu_\mathrm{_I}}
\newcommand{\nuL}{\nu_\mathrm{x}}
\newcommand{\nuV}{\nu_\mathrm{h}}
\newcommand{\nuT}{\nu_\mathrm{t}}
\newcommand{\vsat}{v_\text{sat}}
\newcommand{\Wsat}{W_\text{sat}}
\newcommand{\pc}{p_\mathrm{c}}

\newcommand{\upd}{\mathop{}\!\mathrm{d}}
\begin{document}

\title{Scaling properties of a ferromagnetic thin film model at the depinning transition}

\author[{ }]{M. F. Torres\thanks{mtorres@ifimar-conicet.gob.ar}}
\author[{ }]{R. C. Buceta\thanks{rbuceta@mdp.edu.ar}}
%\author[1,2]{}

\affil[{ }]{Departamento de F\'{\i}sica, FCEyN, Universidad Nacional de Mar del Plata}
\affil[{ }]{Instituto de Investigaciones F\'{\i}sicas de Mar del Plata, UNMdP and CONICET}
\affil[{ }]{Funes 3350, B7602AYL Mar del Plata, Argentina}
%\date{Received: date / Revised version: date}

\maketitle

\abstract{In this paper, we perform a detailed study of the scaling properties of a ferromagnetic thin film model. Recently, interest has increased in the scaling properties of the magnetic domain wall (MDW) motion in disordered media when an external driving field is present. We consider a (1+1)-dimensional model, based on evolution rules, able to describe the MDW avalanches. The global interface width of this model shows Family-Vicsek scaling with roughness exponent $\zeta\simeq 1.585$ and growth exponent $\beta\simeq 0.975$. In contrast, this model shows scaling anomalies in the interface local properties characteristic of other systems with depinning transition of the MDW,  \textsl{e.g.} quenched Edwards-Wilkinson (QEW) equation and random-field Ising model (RFIM) with driving. We show that, at the depinning transition, the saturated average velocity $\vsat\sim f^\theta$ vanished very slowly (with $\theta\simeq 0.037$) when the reduced force $f=p/\pc-1\to 0^{+}$. The simulation results show that this model verifies all accepted scaling relations which relate the global exponents and the correlation length (or time) exponents, valid in systems with depinning transition. Using the interface tilting method, we show that the model, close to the depinning transition, exhibits a nonlinearity  similar to the one included in the Kardar-Parisi-Zhang (KPZ) equation. The nonlinear coefficient $\lambda\sim f^{-\phi}$ with $\phi\simeq -1.118$, which implies that $\lambda\to 0$ as the depinning transition is approached, a similar qualitatively behaviour to the driven RFIM. We conclude this work by discussing the main features of the model and the prospects opened by it.}

\section{Introduction}
The jerky motion of magnetic domain walls (MDW) in ferromagnetic media, when an external driving field is present, is basically the resultant of magnetic interactions, the internal disorder of the medium and the thermal fluctuations. The quenched disorder (or Barkhausen noise) in the media is a consequence of vacancies, defects, impurities and dislocations, among others. The interaction between the molecular magnetic field and the magnetic dipoles locally governs the movement of the MDW between multi-metastable states of quiescence. The dominant dipole-dipole interactions -dipolar (or long-range) and exchange (or short-range)- and the thermal fluctuations favour the elastic displacements of the wall, which contributes to moving it from the quiescence state. The jerky movement of the MDW between quiescence states is known as Barkhausen avalanches or jumps \cite{Barkhausen-19}. This complex microscopic dynamic leads to unexpected properties observed both experimentally and theoretically. 

The study of ferromagnetic films with Barkhausen noise has reopened the study of general systems with avalanches and the discussion about universality classes currently accepted. Such systems can be classified according to the values of the scaling exponents obtained from experiments and models. Systems with magnetic avalanches are classified by measuring the power-law exponents \cite{Durin-05} of the Barkhausen distributions of avalanche-size $\tau$ and -duration $\alpha$, and the average of avalanche-size as a function of the avalanche-duration $1/\alpha\nu z$, and/or the power spectrum $\vartheta$. The studies have led us to accept that there are two universality classes \cite{Durin-05, Sethna-05, Colaiori-08, Mughal-10, Buceta-11}  associated to the range of the dominant interactions (dipolar or exchange), which prevail in amorphous and polycrystalline materials, respectively. For materials that exhibit three-dimensional magnetic behaviour (bulk materials including ribbons and sheets) these two universality classes have been accepted, even when the experimental values of the exponents were widely dispersed within each class. From the beginning, the experimental study of ferromagnetic films has shown dissimilar results according to the techniques employed. The magneto-optical techniques have led to the conclusion that when decreasing the film thickness, the two universality classes are maintained even though the range of exponent values change significantly. The results based on magneto-optical measurements (restricted to distributions of avalanche-sizes), cross-referenced to theoretical predictions, confirm a two-dimensional magnetic behaviour of the thin films \cite{Puppin-00, Kim-03a, Kim-03b, Shin-07, Puppin-07, Ryu-07, Shin-08, Atiq-10, Lee-11a, Lee-11b}; these results are independent of the range of film thickness. This conclusion is consistent with the fully accepted assertion that systems which exhibit universality are independent of the dynamic microscopic details and are controlled by dimensionality and range of interactions, among others properties. Recent reports, based on measurements \textsl{via} the inductive technique, suggest that the two-dimensional behaviour accepted for films cannot be generalized for all thickness ranges \cite{Santi-06, Papanikolaou-11, Bohn-13, Bohn-14}. Some of these experimental results for the scaling exponents grouped films with different thickness in the three-dimensional universality class. In the light of this existing disagreement, we think it is necessary to add to the traditional view another approach followed in the study of interfaces in disordered media. Thus, for ferromagnetic systems with avalanche dynamics there will be at least two criteria for the classification of universality. We focus on the study of the scaling properties of the MDW in a disordered medium through a ferromagnetic thin film model. We establish the exponents characterizing the pinning-depinning transition and their scaling relations.

Driven interfaces in disordered media at the depinning transition can be described by stochastic equations of motion or lattice growth models. Among the local equations, the quenched Kardar-Parisi-Zhang equation (QKPZ) describes systems that can be grouped into two universality classes for their properties at the depinning transition: isotropic and anisotropic. The two classes differ fundamentally in the  coefficient behaviour of the nonlinearity near the transition, as shown below. Some lattice models that can be identified with one of these two classes of universality, \textsl{e.g.} random field Ising model with driving belongs to the isotropic universality class \cite{Cieplak-88, Martys-91} while directed percolation depining (DPD) models belong to the anisotropic universality class \cite{Tang-95, Albert-98}. There are other models which cannot be included in a QKPZ description type, such as the interface evolution models in fractal media, which are grouped in the called isotropic percolation depinning (IPD) universality class \cite{Asikainen-02}. The QKPZ equation for the height $h=h(x,t)$ is
\begin{equation}
\frac{\partial h}{\partial t}=F+\nu\,\nabla^2 h+\frac{\lambda}{2}\abs{\nabla h}^2+\eta(x,h)\;,\label{QKPZ}
\end{equation} 
where $F$ is an external driving force and $\eta$ is the quenched noise with mean value equal to zero and correlation $\langle\eta(x,h)\,\eta(x',h')\rangle=2\,D\,\delta(x-x')\,\Delta(h-h')$. Here $\Delta$ is an even function which decreases monotonically, decaying rapidly to zero at a characteristic distance $a$, and $\Delta=\delta$ if $a=0$. In the anisotropic universality class at the depinning transition the nonlinear coefficient $\lambda$ diverges when the reduced force $f\to 0^+$. In contrast, the isotropic class $\lambda\to 0$ when the $f\to 0^+$. Taking $\lambda=0$ in equation~(\ref{QKPZ}), the quenched Edwards-Wilkinson (QEW) equation is achieved, and the isotropic class is usually called QEW universality class. 

There are two possible theoretical approaches to explain the depinning transition of the driven MDW in a disordered medium: the QEW equation \cite{Duemmer-05,Kolton-06a, Kolton-06b, Bustingorry-08} and the microscopic lattice models with interactions and disorder, {\sl e.g.} driven RFIM \cite{Qin-12,Zhou-09,Zhou-10,Roters-01}. The QEW equation is a phenomenological coarse-grained equation which does not take into account interactions and structure. The MDW described by the QEW equation is a single-valued elastic surface \cite{Kolton-05, Braun-05,Kleemann-07}. In contrast, the MDW described by the known lattice models is not single-valued as a result of islands and/or overhangs \cite{Urbach-95,Roters-00}. Equations and models can be classified according to their scaling properties (usual and anomalous), taking into account the relationship between the global, local and spectral roughness exponents \cite{Lopez-97c,Lopez-99,Ramasco-00}. The QEW equation and driven RFIM belong to the anomalous scaling class. The subclass of anomalous scaling (intrinsic, super-rough, faceted, or other) to which these systems belong is currently still under study \cite{Qin-12,Zhou-09,Zhou-10,Chen-10,Rosso-03, Pradas-08, Pang-00,Rosso-01,Jost-97,Jost-98}. 

In a recent paper Muraca and Buceta (BM) have introduced an evolution model of the MDW that takes into account only short-range interactions, in order to describe avalanches in a ferromagnetic thin film with disorder when a driving field is present \cite{Buceta-11}. They have found that at the depinning transition the model predicts $\tau = 1.29\pm 0.02$ and $\alpha = 1.55\pm 0.05$, which are similar to the results obtained experimentally in films where exchange interactions dominate \cite{Bohn-14, Kim-03b, Ryu-07, Shin-08}. In our previous work \cite{Torres-13} we showed that the BM model has anomalous scaling in its local statistical observables, similarly to the QEW equation and the driven RFIM but with different anomalous class. The BM model has intrinsic anomalous scaling, while the QEW equation has super-rough anomalous scaling \cite{Lopez-97a} and the driven RFIM does not have an anomalous scaling univocally defined \cite{Chen-10}. Given these interesting properties we decided to study the global scaling at the depinning transition using a simplified version of the BM model introduced by us, which maintains all the properties listed above. In Section 2, we present the lattice model under study. In Section 3, we present the theoretical framework that supports the results obtained from the Monte Carlo simulations performed, and establish scaling relations between the exponents of the global roughness and average velocity associated with the depinning transition. In Section 3, we also analyse the simulation results and show how these results are in agreement with the theory. In Section 4, we show how our model, at the depinning transition, behaves similarly to other systems that are within the isotropic universality class. Using the method of the tilted surface, the simulation results show that near the transition there is a QKPZ nonlinearity, which vanishes at criticality. Finally, a summary is given in Section 5. 

\section{The model\label{sec:model}}

We consider a portion of the ferromagnetic thin-film that includes only two magnetic domains separated by a MDW. On either side of the wall we assume opposite macroscopic magnetizations in the easy direction. We suppose that the medium is composed of magnetic dipoles and point defects randomly distributed, both arranged in the nodes of a 2-dimensional square lattice of edge length $L$. The BM model assumes that two defects cannot be first-neighbours to each other. In contrast, in this model we assume that defects are arranged in a completely random way without restrictions. The wall in its movement does not include defects. The structure of the MDW is considered merely as a monolayer of dipoles perpendicular to the easy direction. The model takes into account only exchange interactions between nearest-neighbour (NN) dipoles. To simulate the movement of the MDW, we assume that the lattice has periodic boundary conditions and that point defects can be represented by a random pinning force $\eta(i,j)$, uniformly distributed in $[0, 1]$, assigned to each node $(i,j)$ of the lattice, with $i=1,\dots,L$ and $j\in\mathbb{N}$. Taking a lattice with density $p$ of dipoles and \mbox{$1-p$} of defects, if $\eta(i, j) < p$ the node $(i, j)$ has a dipole; otherwise, it has a point defect. We characterize the disorder in the lattice using the function $F (i, j) = \Theta (p - \eta(i, j))$, where $\Theta(x) = 1$ if $x\ge 0$ and $\Theta(x) = 0$ if $x < 0$. If $(k, n_k)$ is a node of the MDW with $k = 1,\dots , L$, as the wall is a monolayer of dipoles $F(k,n_k)=1$. Otherwise, any dipole (or defect) outside the MDW is described by $F (k,\ell) = 1\,(\text{or}\,0)$ with $\ell\neq n_k$.

The evolution rules of this model consider only ferromagnetic exchange (or short range) interactions, taking into account the balance of the magnetic moment on either side of the MDW in a neighbourhood of the point which is evolving. When there is a local unbalance in the opposite direction of the movement, the wall searches the equilibrium with probability $c$. However, if there is balance, with probability $1 -c$, the wall can only move if there is external forcing which can remove it from its metastable state. We start the Monte Carlo simulation with a flat wall, {\sl i.e.} initial condition $n_i = 1$ for all $i$. We introduce, at a fixed time on the MDW, the relative position of neighbouring nodes respect to the node of the selected column:
\begin{equation}
x_j = n_{j+1} - n_j\;,\hspace{5ex} y_j = n_{j-1} - n_j\;.\label{xyj}
\end{equation}
At time $t$ any column $j$ is chosen randomly. The MDW will advance from the selected node $(j, n_j)$ to the node $(j,n_j')$, in the interval $\delta t$, according to simple evolution rules. Those rules are frustrated if $F(j,n_j')=0$ or $F(j,n_j'-1)=0$; in the first case because the MDW is composed only by dipoles, and in the second case because the MDW needs a larger unbalance of energy to overcome a defect. The evolution rules are
\begin{itemize}
\item[1.] With probability $c$, if $x_j + y_j \ge 2$ the selected node is moved one unit ({\sl i.e.} $n_j' = n_j + 1$); otherwise the selected node is pinned ({\sl i.e.} $n_j' = n_j$). 
\item[2.] With probability $1-c$, if $x_j = y_j \ge 0$ the selected node is moved one unit above its NN columns ({\sl i.e.} $n_j' = n_{j+1} + 1$); otherwise the selected node moves (or not) to the maximum between the selected column and its NN columns ({\sl i.e}. $n_j'=\max(n_{j-1},n_j,n_{j+1})$). 
\end{itemize}
These rules applied over a lattice with randomly distributed defects lead to the same properties as those obtained by the BM model: avalanches and scaling anomalies, with similar quantitative conclusions \cite{Buceta-11, Torres-13}. 

\section{Scaling properties\label{sec:scaling}}

\begin{table}
\begin{center}
\begin{tabular}{|c|c|c|c|c|c|c|}
\hline 
$p^*$ & $\theta$ & $\nuV$ & $\beta$ & $\beta_\mathrm{c}$ \\ 
\hline 
0.88700 & 0.0415(5) & 1.6206(151) &  1.0228(11) & 0.9750(5)\\ 
\hline 
0.88710 & 0.0389(6) & 1.5224(116) &  0.1005(12) & 0.9751(5)\\ 
\hline 
0.88715 & 0.0376(6) & 1.4730(100) &  0.9777(1) & 0.9751(5)\\ 
\hline 
0.88716 & 0.0374(6) & 1.4631(97) &  0.9749(2) & 0.9751(5)\\ 
\hline 
0.88717 & 0.0371(6) & 1.4532(94) &  0.9699(4) & 0.9751(5)\\ 
\hline 
0.88720 & 0.0364(6) & 1.4234(85) &  0.9565(5) & 0.9751(6)\\ 
\hline
0.88730 & 0.0338(7) & 1.3233(65) &  0.9195(12) & 0.9751(6)\\ 
\hline 
\end{tabular} 
\caption{We show the measured values of the power-law exponents (columns 2-4) as a function of the control parameter $p^*$ very close to the depinning transition for a lattice size $L=10^4$. Columns 2 and 3 show the exponents $\theta$ and $\nuV$ of the $\vsat$ and $\Wsat$, respectively, as a function of reduced force $f=p/p_\mathrm{c}-1$, with  $\pc\simeq p^*$, measured in the interval $[10^4,3\times 10^4]$. Column 4 shows the exponent $\beta$ of the $W$ as a function of time $t$ measured in the same interval. Column 5 shows the growth exponent $\beta_\mathrm{c}$ calculated by equations~(\ref{scaling-1}) and (\ref{scaling-2}) with the exponent data $\theta$ and $\nuV$ (columns 2-3). Notice that the best approximation to the critical value $\pc$ is obtained for $L=10^4$ with $p^*=0.88716$.\label{table1}}
\end{center}
\end{table}
\begin{figure}
\begin{center}
\includegraphics[scale=.3]{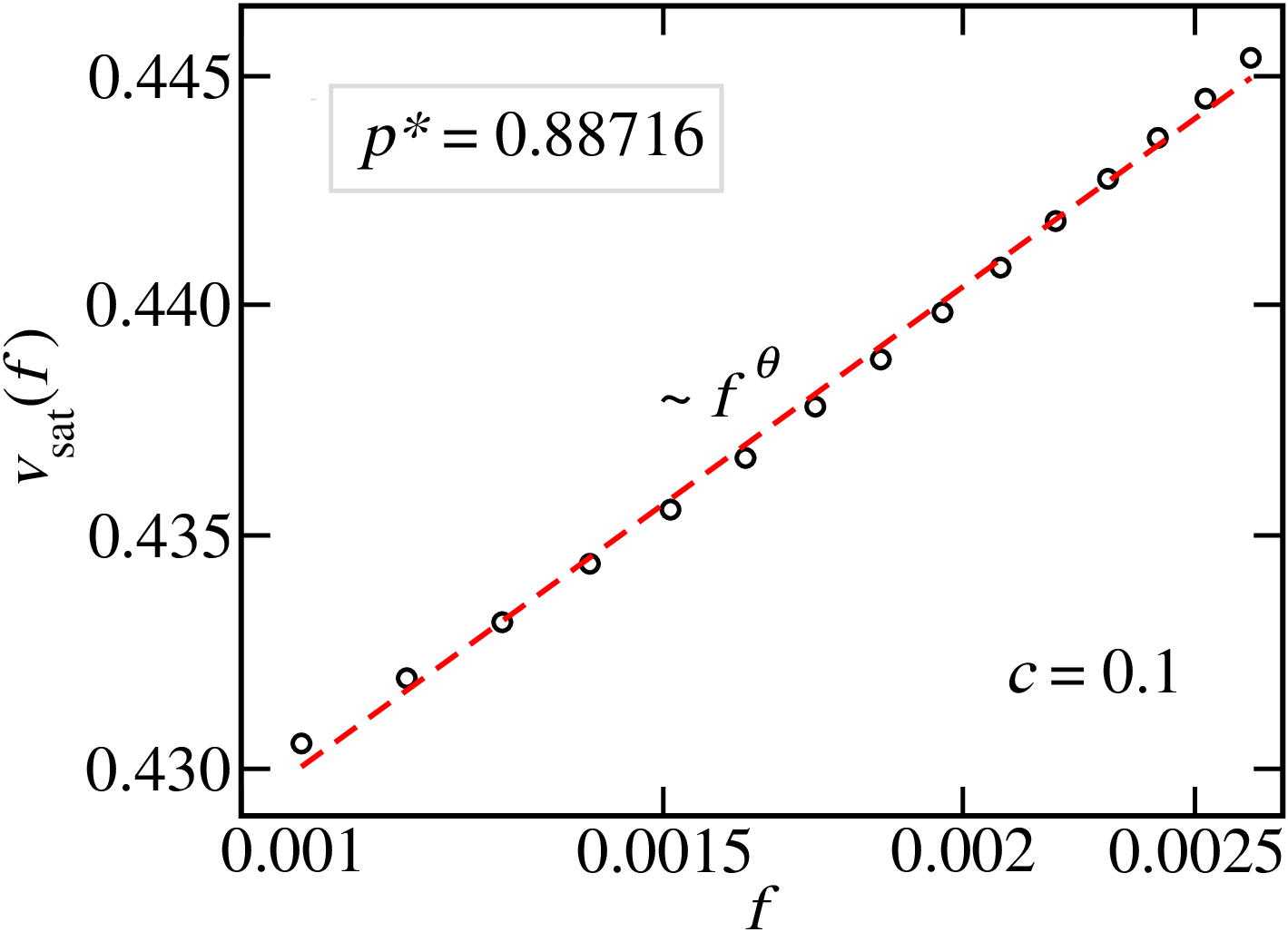}\hspace{1.2cm}
\includegraphics[scale=.3]{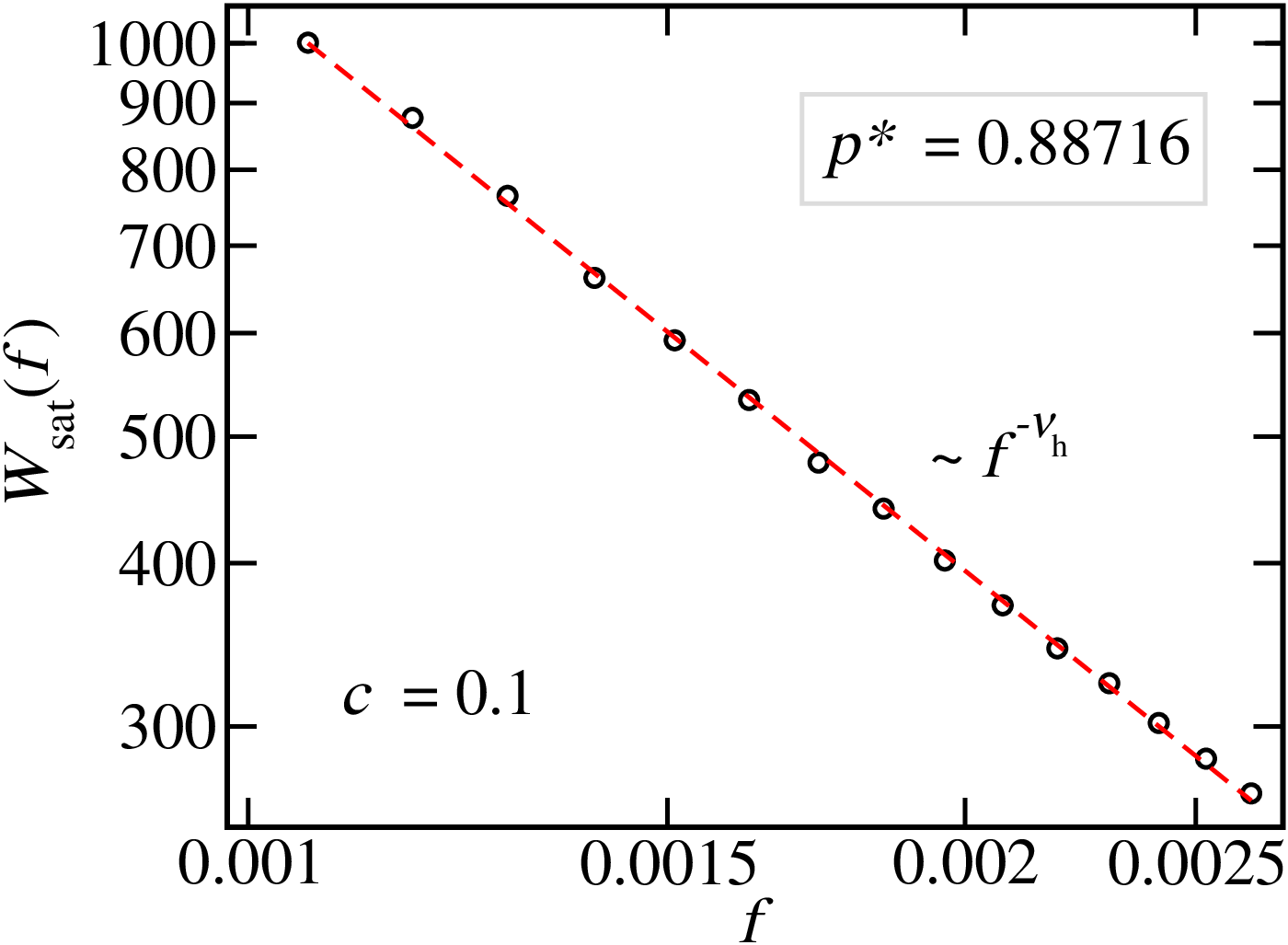}
\end{center}
\caption{(Color online) Left: log-log plot of the average saturation velocity $\vsat$ as a function of the reduced force $f=p/\pc-1$ close to the depinning transition, with $\pc\simeq p^*=0.88716$. The measured exponent is $\theta=0.0376(6)$ (see Table~\ref{table1}). Right: log-log plot of the saturation global interface width $\Wsat$ as a function of the reduced force $f$ close to the depinning transition. The measured exponent is $\nuV=1.4631(97)$ (see Table~\ref{table1}). We take the parameters $p>\pc$ and $c=0.1$, and lattice size $L=10^4$ in the simulations corresponding to both graphs.\label{fig:1}}
\end{figure}
The scaling theory applied to driven interfaces in disordered media has developed strongly in the last three decades, starting from models and equations to describe the phenomenon of pinning-depinning transition \cite{Barabasi-95}. Such phenomenon is observed in the MDW motion \cite{Metaxas-07}, the imbibition of fluids in porous media \cite{Pradas-09} and flux motion in superconductors \cite{Fisher-91}, among others. The theoretical studies are based on the average velocity of the interface, which plays the role of order parameter of the pinning-depinning transition. We assume that the system enables a control parameter $p$, which takes a critical value $\pc$, such that if $p<\pc$ the interface becomes pinned by disorder and, otherwise it is able to move indefinitely with an average velocity $v(f)$, where $f=p/\pc-1$ is the reduced force. When $p\to\pc^{+}$, the average saturation velocity $\vsat$ vanishes as 
\begin{equation}
\vsat(f)\sim f^{\theta}\,,\label{exp1}
\end{equation}
where $\theta>0$. Near the criticality $\pc$, large and finite regions of the interface are pinned by disorder. At the transition, there are correlation lengths (and correlation time) $\xiI$ of these pinned regions that diverge as
\begin{equation}
\xiI\sim \abs{f}^{-\nuI}\,,\label{exp2}
\end{equation}
where $\nuI>0$ is a correlation length (or time) exponent.  Here $\xiI$ refers to the lateral (or parallel) correlation length $\xiL$, the vertical (or perpendicular) correlation length $\xiV$, or the correlation time $\xiT$. At the criticality $p=\pc$, in a finite system with lateral size $L$, the dynamic scaling hypothesis is that the global interface width $W$ verifies the Family-Vicsek scaling relation \cite{Family-85} 
\begin{equation}
W(L,t)\sim L^\zeta\;F(t/L^z)\,,\label{giw}
\end{equation}
where $\zeta$ and $z$ are the roughness and dynamic exponents, respectively. The scaling function $F(x)\sim x^\beta$ when $x\ll 1$ and goes to a constant when $x\gg 1$, where the growth exponent $\beta$ fulfils the relation
\begin{equation}
\zeta=\beta\,z\;.\label{FVscaling}
\end{equation}
Also, at the criticality for a large and finite system, the average velocity verifies the scaling relation 
\begin{equation}
v(L,t)\sim L^{-z\delta}\,G(t/L^z)\,,\label{av}
\end{equation} 
where $\delta$ is the velocity exponent. The scaling function $G(x)\sim x^{-\delta}$ when $x\ll 1$ and goes to zero when $x\gg 1$. The exponents that characterize the pinning-depinning transition [by eqs.~(\ref{exp1}) and (\ref{exp2})] are related to the exponents $\delta$\,, $\zeta$ and $z$ in a simple way, as we show below.

Near the depinning transition, large regions of the interface are pinning. The full pinning is achieved when the lateral correlation length $\xiL$ is of the order of the system size $L$ ({\sl i.e.} $\xiL\sim L$), the vertical correlation length $\xiV$ is of the order of the global interface width $W$ ({\sl i.e.} $\xiV\sim W$) and the correlation time $\xiT$ is of the order of the crossover time $t_\textsf{x}$ ({\sl i.e.} $\xiT\sim t_\textsf{x}$). Here, the crossover time $t_\textsf{x}$ is the time at which the correlations are propagated throughout the system, saturating the global interface width, thus $t_\textsf{x}\sim L^z$. Eq.~(\ref{exp2}) shows that $\xiV\sim\xiL^{\nuV/\nuL}$, which allows us to obtain that $W\sim L^{\nuV/\nuL}$. Taking into account that $W\sim L^\zeta$\; when $t\simeq t_\textsf{x}$ [see eq.~(\ref{giw})], we conclude that $\nuV =\zeta\;\nuL$. Also eq.~(\ref{exp2}) shows that  $\xiV\sim\xiT^{\nuV/\nuT}$, which allows us to obtain that $W\sim t_\textsf{x}^{\nuV/\nuT}$. Considering that $W\sim t_\textsf{x}^\beta$\; when $t\simeq t_\textsf{x}$ we conclude that $\nuV =\beta\;\nuT\,$. Thus, we obtain the following relations between correlation exponents
\begin{equation}
\nuV =\zeta\;\nuL =\beta\;\nuT\;.\label{scaling-1}
\end{equation}
Close to the criticality $\pc$ but below it, the average velocity of the interface $v\sim t_\textsf{x}^{-\delta}$ when $t\simeq t_\textsf{x}$, and later the velocity goes to zero quickly [see eq.~(\ref{av})]. In terms of the reduced force $f$ the velocity $v\sim\abs{f}^{\delta\nuT}$. Taking into account eq.~(\ref{exp1}) we obtain the relation $\theta=\delta\,\nuT\,$ if $f>0$ \cite{Hinrichsen-00}. Assuming that the stationary average velocity $v_\text{st}\sim\xiV/\xiT$ it is easy to conclude that $v_\text{st}\sim\abs{f}^{\nuT-\nuV}$ and $\theta=\nuT-\nuV\,$ if $f>0$. Thus, we obtain the following additional relations between correlation exponents
\begin{equation}
\theta=\nuT-\nuV=\delta\,\nuT\;.\label{scaling-2}
\end{equation}  
\begin{figure}[ht]
\begin{center}
\includegraphics[scale=.3,height=5cm]{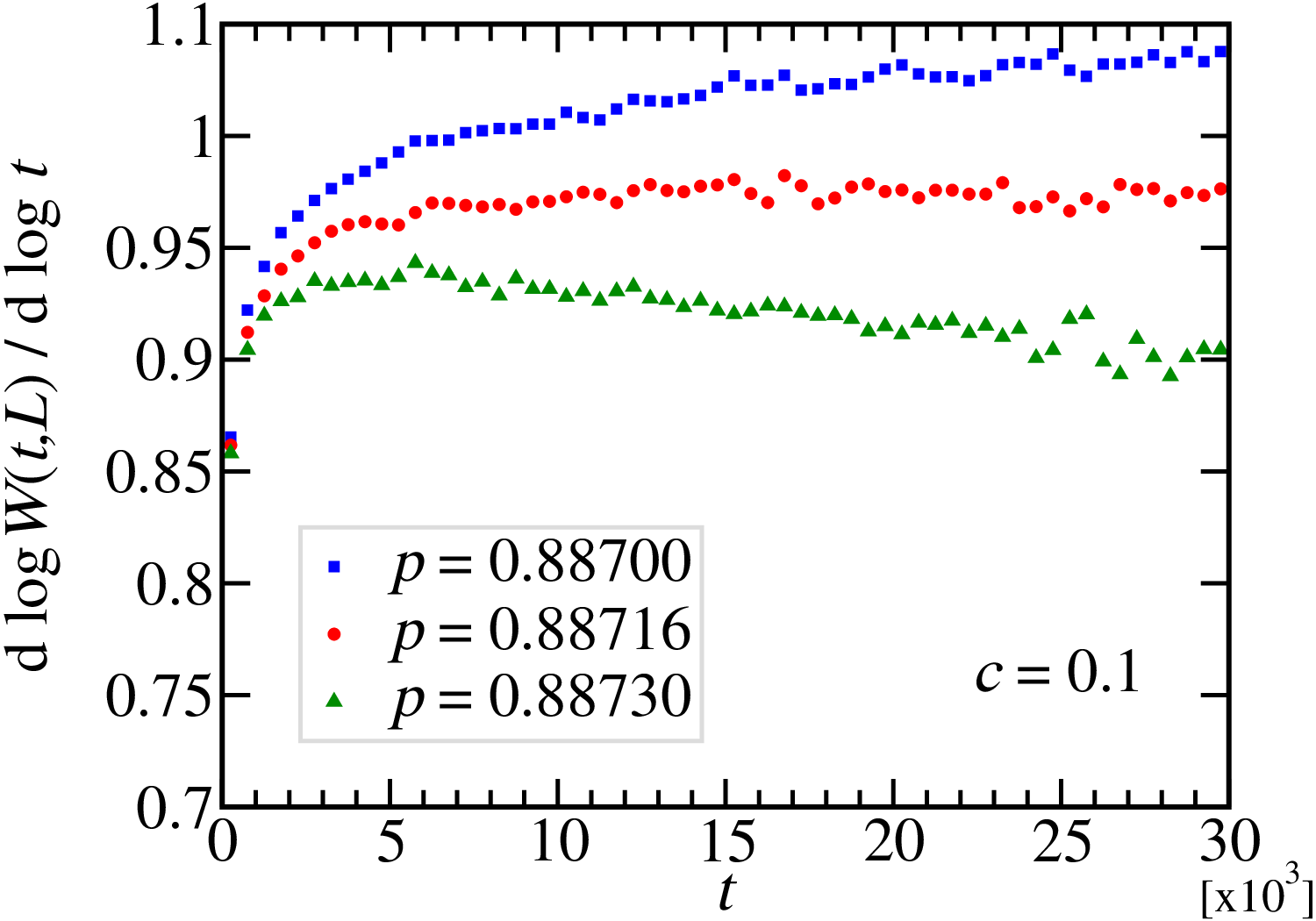}
\hspace{1.2cm}\vspace{1ex}
\includegraphics[scale=.3,height=5.2cm]{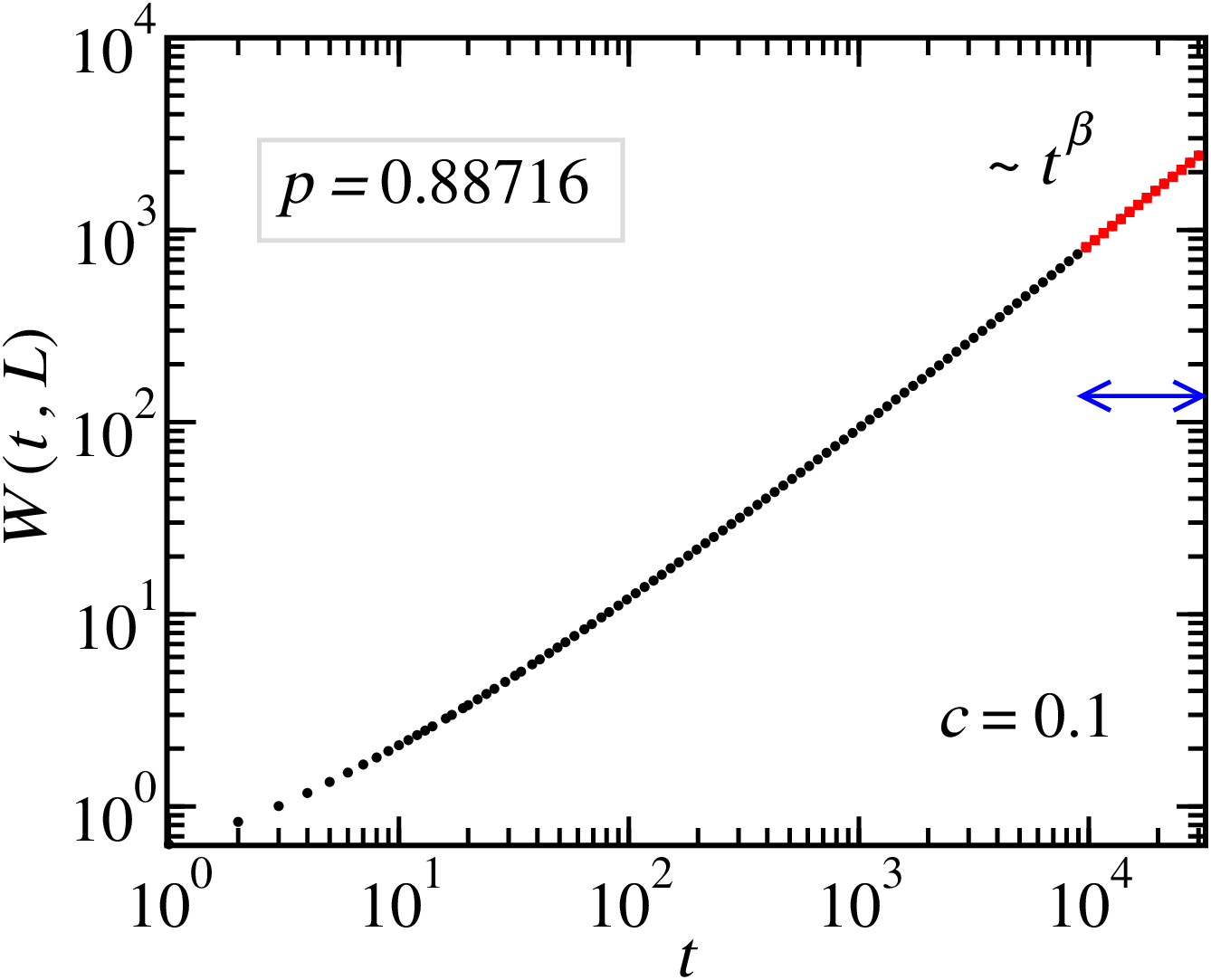}
\end{center}
\caption{(Color online) Left: plot of $\frac{\upd\log W}{\upd\log t}$ as a function of time $t$. We show the simulation results for $c=0.10$ and $L=10^4$ and several values of $p$ very close to critical value. Notice that for $p=0.88716$, very close to the depinning transition, the curve approaches a constant in the interval $[10^4,3\times 10^4]$. Right: log-log plot of $W$ as a function of time $t$. The double arrow $\leftrightarrow$ shows the interval where the growth exponent $\beta=0.9749(2)$ is measured.
\label{fig:2}}
\vspace{1ex}
\end{figure}
\begin{figure}
\begin{center}
\includegraphics[scale=.3]{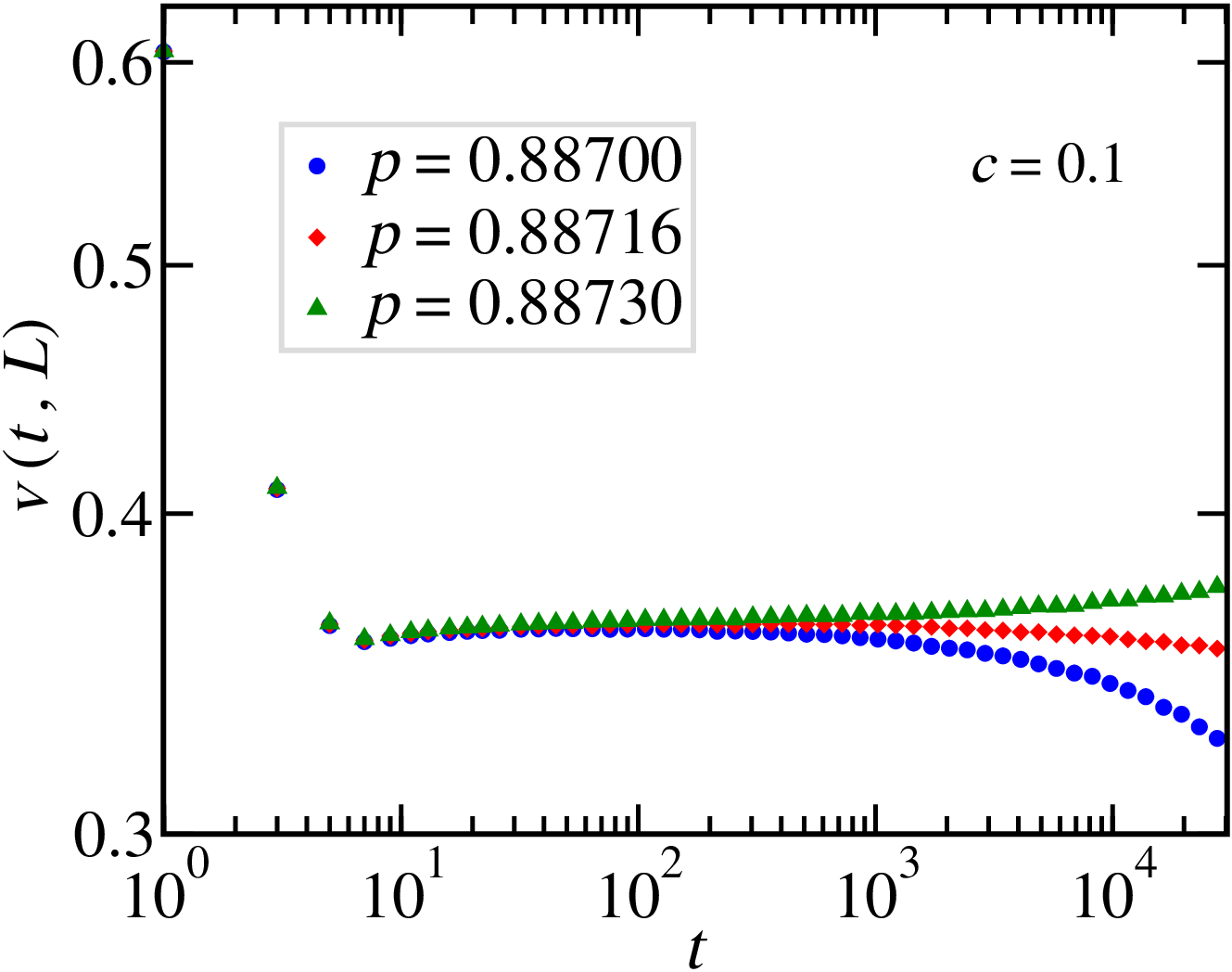}
\end{center}
\caption{(Color online) Log-log plot of the velocity $v$ as a function of time $t$ with lattice size $L=10^4$ and $c=0.10$, for several values of $p$ arround the critical value. When $p=0.88716$ (red diamond symbols) very close to criticality we observe the power law behaviour $v\sim t^{-\delta}$, with $\delta=0.0113(62)$ which is measured in the interval $[10^4,3\times 10^4]$. \label{fig:3}}
\end{figure}
\begin{figure}[ht]
\vspace{1ex}
\begin{center}
\includegraphics[scale=.3]{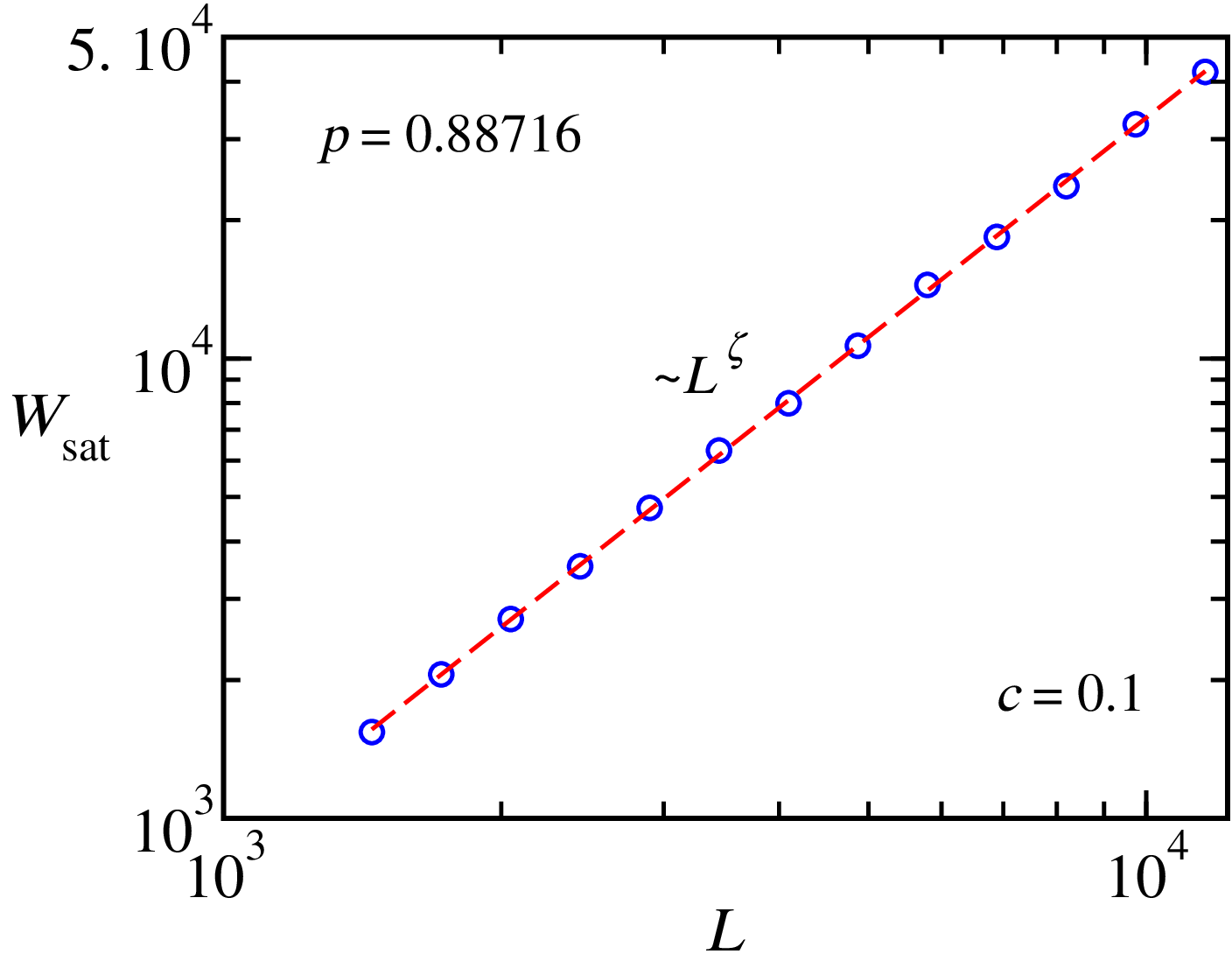}
\end{center}
\caption{(Color online) Log-log plot of the saturation global interface width $\Wsat$ as a function of lattice size $L$ with $p=0.88716$ close  to criticality and $c=0.10$. We measure the power law behaviour $\Wsat\sim L^{\zeta}$ with $\zeta=1.5851(68)$.\label{fig:4}}
\vspace{3.5ex}
\end{figure}
Intensive Monte Carlo simulations of our model with large systems close to the depinning transition indicate that the growth exponent is $\beta\lessapprox 1$, which is confirmed through the calculus of the power-law exponent of the global interface width $W$ and the saturation height-height correlation function $G_{2,\text{sat}}$\,\cite{Torres-13}. To set $\beta$ as accurately as possible it is first necessary to establish the value of $p$ which is closest to its value at the depinning transition. A usual procedure is to determine the critical value $p_\mathrm{c}$ by epidemic analysis \cite{Grassberger-79, Grassberger-89} of the velocity as a function of time. At criticality if $t\ll L^z$ the velocity $v\sim t^{-\delta}$. From equations~(\ref{scaling-1}) and (\ref{scaling-2}) we can deduce that velocity and growth exponents are related by
\begin{equation}
\delta+\beta=1\;,\label{scaling-3}
\end{equation}
equation that is strictly valid in the thermodynamic limit at the depinning transition. By accepting this equation we deduce that $\delta\gtrapprox 0\,$ for our system, which makes the application of the epidemic analysis method a disadvantage. Alternatively, the suitable power-law behaviour of the saturation velocity $\vsat$ and the saturated interface width $\Wsat$ as a function of the reduced force $f$ provides a method to establish the critical value $\pc$. Table~\ref{table1} shows the measured exponents $\theta$ of $\vsat$ and $\nuV$ of $\Wsat$ for several values of the control parameter $p^*$ close to the depinning transition. With these data we calculate the growth exponent $\beta_\mathrm{c}\simeq 0.9751$ for different values of $p^*$. In contrast, the direct measurement of $\beta$ shows that the value $\beta_\mathrm{c}$ is reached for $p^*=0.88716$ very close to the depinning transition, with $\beta\simeq 0.9749$. Figure~\ref{fig:1} shows the power-law behaviour of $\vsat$ and $\Wsat$ as a function of $f$ very close to the critical value $p_\mathrm{c}$. Figure~\ref{fig:2} shows the interval where the critical power-law behaviour of $W$ as a function of time $t$ is measured. Figure~\ref{fig:3} shows the power-law behaviour of the velocity as a function of time, with measured exponent $\delta=0.0113(62)$ for $L=10^4$, close to the transition. With this method, we find $\delta+\beta=0.9862(64)$ for the same lattice size, a very good approximation to the equation~(\ref{scaling-3}). We determine $\zeta= 1.5851(68)$ from the saturation interface  width $\Wsat$ as a function of size $L$ for $p=0.88716$ (see Figure~\ref{fig:4}). With the measured values from Table~\ref{table1} we obtain, by equation~(\ref{scaling-1}), $\nuT= 1.5008(103)$ and $\nuL=0.9230(100)$. 

The fact that $\beta\lessapprox 1$ (or $\theta\gtrapprox 0$) has important consequences. The validity of equation (\ref{exp1}) is confirmed because $\theta$ is never zero. Thus, the interface velocity vanishes very slowly when the system approaches criticality. This hallmark places our model in the limit of the known systems with depinning transition. Notice that as $\nuT\approx\nuV\,$; then, at the transition, the vertical correlation length $\xiV$ and the correlation time $\xiT$ diverge similarly. In contrast, the DPD models verify that $\nuT\approx\nuL$ [from eq.~(\ref{scaling-1}) with $\zeta\approx\beta$]; then, at the transition, the lateral correlation length $\xiL$ and the correlation time $\xiT$ diverge similarly.

\section{Behaviour at the depinning transition}

The method of measuring the saturation velocity of the tilted interface, for different values ​​of the reduced force $f$, is useful not only to determine QKPZ nonlinearities but also to establish the universality class at the depinning transition. For a fixed value of $f$ when the interface velocity depends on the slope $s$ the system includes nonlinearities. The coarse-grain limit only maintains quadratic nonlinearities, which are evident with slopes $s\ll 1$. The average saturation velocity of the tilted interface is
\begin{equation}
\vsat(s,f)=\vsat(f)+\lambda(f)\,s^2\;,\label{exp3}
\end{equation}
where $\vsat(f)$ is the saturation velocity of the untilted surface [eq.~(\ref{exp1})] and $\lambda(f)$ is the coarse-grained nonlinear coefficient. The velocity of driven interfaces in disordered media, near the depinning transition ($f\to 0^+$), shows two opposite behaviours. In systems included within the anisotropic universality class $\lambda$ diverges when $f$ goes to zero. In contrast, other systems are included within the isotropic universality class at the depinnig transition when $\lambda$ approaches zero. Simulations of several models show that
\begin{equation}
\lambda\sim f^{-\phi}\;.\label{exp4}
\end{equation}
At the depinning transition, if the tilt exponent $\phi>0$ ($\phi<0$) then the coefficient $\lambda$ diverges (vanishes) and the model belongs to the anisotropic (isotropic) universality class. Substituting equations~(\ref{exp1}) and (\ref{exp4}) into equation~(\ref{exp3}) we find \cite{Amaral-94}
\begin{equation}
\vsat(s,f)\propto f^{\theta}+\gamma\,f^{-\phi}\,s^2\;,
\end{equation}
with $\gamma>0$ a constant and $\theta>0$. Numerical simulation results of different systems show that $\vsat$ is an increasing function of $f$, {\sl i.e.} $\upd\vsat/\upd f\propto \left[\theta-\phi\,\gamma\,s^2\,f^{-(\theta+\phi)}\right] f^{\theta-1}>0$. In systems included within the anisotropic universality class $\phi>0$ and $\theta\simeq\phi$ \cite{Tang-92,Buldyrev-92,Amaral-95}; then, $\vsat$ always increases if $s^2< s^2_\texttt{c}=(\theta/\gamma\,\phi) f^{\theta+\phi}$, where $s_\texttt{c}$ is the cutoff slope. Instead, in systems included within the isotropic universality class (at the depinning transition) $\phi<0$, then $\vsat$ always increases with $f$. Introducing the crossover slope $s_\texttt{x}$ such that $s^2_\texttt{x}\sim f^{\theta+\phi}$, the systems included in these universality classes show the following scaling for the tilted saturation velocity
\begin{figure}
\begin{center}
\includegraphics[scale=.3]{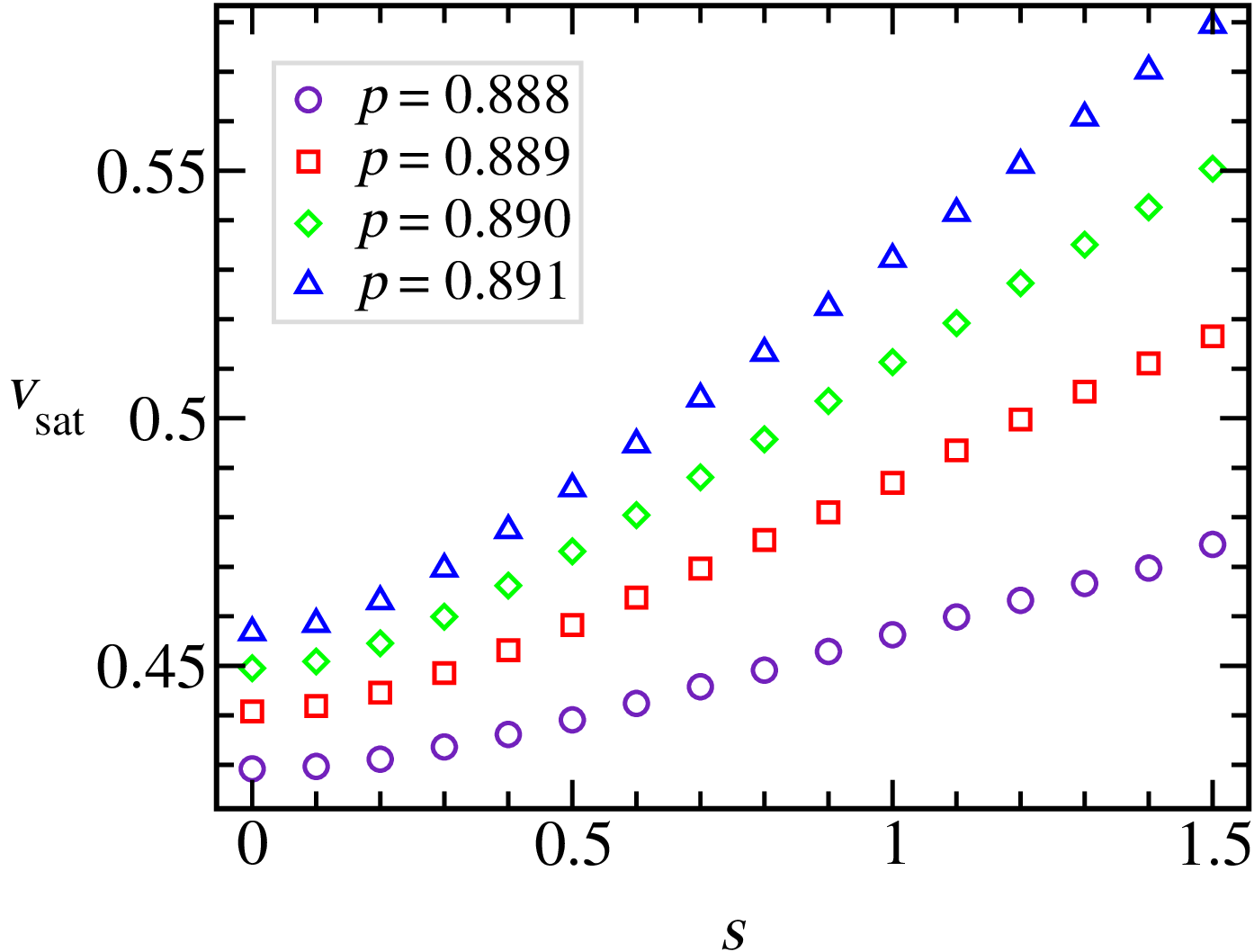}\hspace{1.2cm}
\includegraphics[scale=.3]{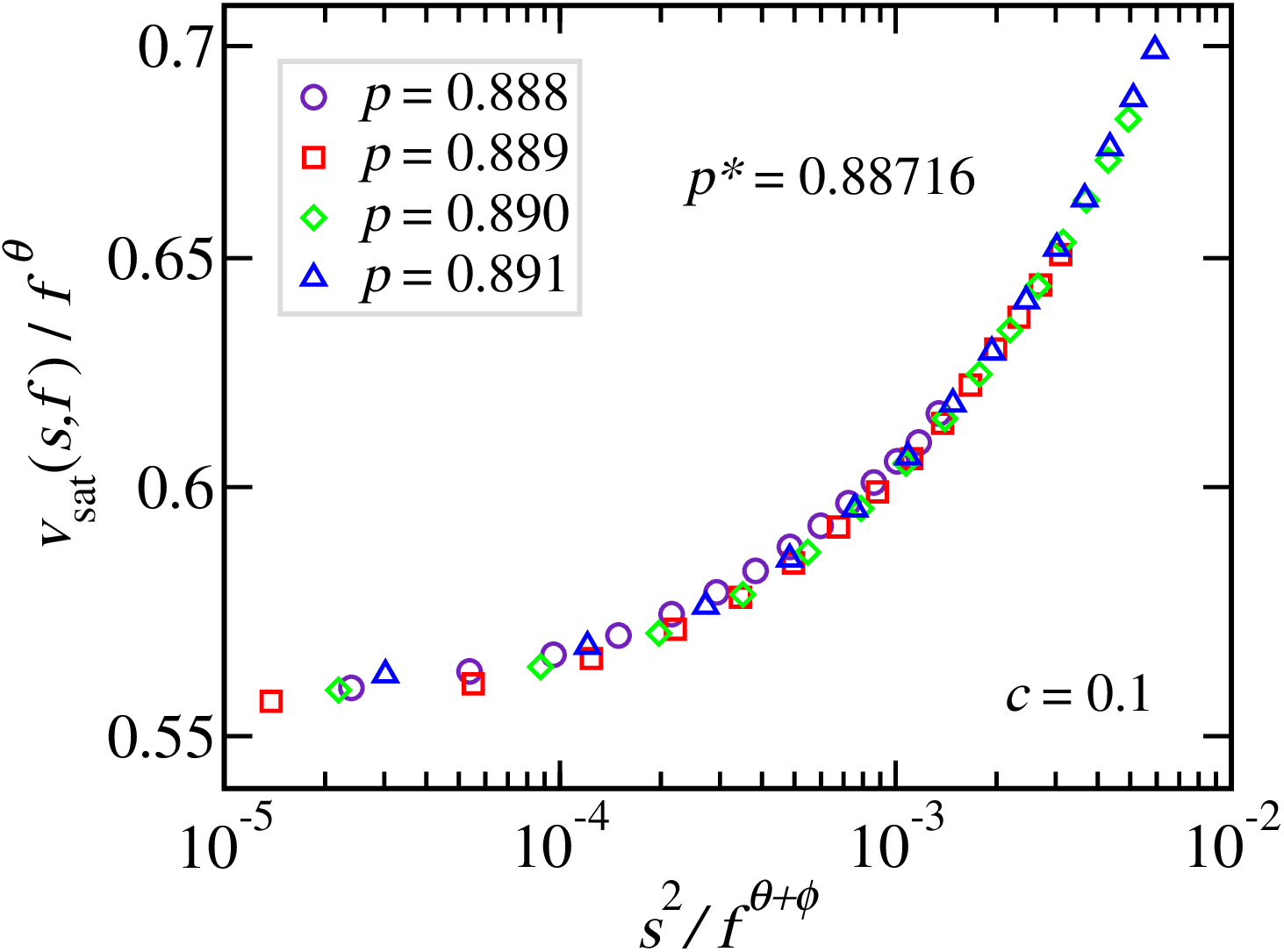}
\end{center}
\caption{(Color online) Left: Plot of $\vsat$ as a function of the tilt $s$ for several values of $p$ above the depinning transition. We note quadratic dependence for $s<1$. Right: log-log plot of the scaling function $V$ as a function of $x$ [see equation~(\ref{tilted-vsat})] for several values of $f=p/\pc-1$, with $\pc\simeq p^*=0.88716$. We take $\theta=0.0376$ and $\theta+\phi=-1.0801$.  
\label{fig:5}}
\end{figure}
\begin{equation}
\vsat(s,f)\sim f^\theta\,V\!\left(\frac{s^2}{f^{\theta+\phi}}\right)\;,\label{tilted-vsat}
\end{equation}
where scaling function $V(x)\sim x$ when $x\gg 1$ and goes to a constant when $x\ll 1$. Introducing the characteristic slope $\tilde{s}=\xiV/\xiL$ it is possible to relate the exponent $\phi$ to other known exponents \cite{Tang-95}. If $\tilde{s}\sim s_\texttt{x}$ [as $\tilde{s}^2\sim\abs{f}^{2(\nuL-\nuV)}$ and $s^2_\texttt{x}\sim f^{\theta+\phi}$]; then, for $f>0$ we conclude that
\begin{equation}
\theta+\phi=2(\nuL-\nuV)\;.\label{exp5}
\end{equation}
Substituting equation~(\ref{scaling-1}) into equation~(\ref{exp5}) we obtain
\begin{equation}
\theta+\phi=2\,\nuV\left(\frac{1}{\zeta}-1\right)\;.
\end{equation}
With the measured values of $\nuV$, $\theta$ and $\zeta$ for $\pc\simeq p^*= 0.88716$ (from Table~\ref{table1}) we obtain $\phi\simeq -1.1177(157)$ [or $\theta+\phi=-1.0801(151)$]. We use these values (see Figure~\ref{fig:5}) to show that the scaling relation [equation~(\ref{tilted-vsat})] for the tilted saturation velocity is verified in our model.  The negative value of $\phi$ means that $\lambda\to 0$ when $f\to 0^+$ [from equation~(\ref{exp4})], a characteristic behaviour of driven RFIM and QEW equation.

\section{Conclusions\label{sec:concl}}

In summary, in this work we perform a detailed study of the scaling properties of the driven MDW in a disordered medium at the depinning transition for a ferromagnetic thin film model. We have chosen this model for the study of scaling since it has proven to be highly predictive. First, the model has demonstrated to be in agreement with the experimental results of the distributions of avalanche-size and -duration. Second, this model has predicted scaling anomalies such as the driven RFIM model and QEW equation. In this paper, we apply the known scaling theory that has been developed to understand the motion of driven interfaces in disordered media. We have found that the driven MDW verifies the scaling relations established, although with some peculiarities. We show that at the depinning transition ($p=\pc$) the relation between scaling exponents $\beta+\delta=1$ holds. Very close to the depinning transition, we show that the growth exponent $\beta\lessapprox 1$, which is a particular feature of this model. As a consequence, the vertical correlation length $\xiV$ and the correlation time $\xiT$ diverge similarly at the depinning transition. Furthermore, the average saturation velocity as a function of time shows a power-law behaviour that decays very slowly, approaching a constant, since the velocity exponent $\delta\gtrapprox 0$. Also, the average saturation velocity of the interface vanishes very slowly when the reduced force goes to zero due to the exponent $\theta\gtrapprox 0$. The values of these exponents differ from the range of values obtained by other authors for driven RFIM and QEW equation \cite{Amaral-95, Kim-06}. Still, we show that this model verifies all relations between scaling exponents that are valid for systems with depinning transition. We conclude that our model, at the depinning transition, behaves similarly to systems that are within the isotropic universality class. Results of simulating our model with the tilted surface show that there is a KPZ nonlinearity, which vanishes at criticality. The tilt exponent determined by the simulations confirms the known relation with other scaling exponents. When our model approaches the depinning transition we observe a similar qualitatively behaviour to the driven RFIM. However, the values of the exponents of the two models differ appreciably. In contrast, the values of the exponents approach each other in the driven RFIM and the QEW equation. We think that this difference in the values of the exponents of our model with others could be given by correlations of quenched disorder in the motion of a MDW. This conjecture is suitable taking into account the results of driven RFIM with different forms and strengths of disorder, which lead to a wide dispersion of the exponent values \cite{Qin-12}. Based on these conclusions, future studies should focus on a more general classification of universality to describe the statistical properties of the MDW moving in disordered media. We believe that this discrete model is a good candidate to be studied further in order to obtain the Langevin equation that characterizes the universality class of the system, by using coarse-grained techniques \cite{Vvedensky-93, Buceta-12}.

\section*{Acknowledgement}
R.C.B. thanks C. Rabini for her suggestions on the final manuscript.

\bibliography{TB-2015}

\end{document}